\documentclass[aip, amsmath, amssymb,]{revtex4-1}

\usepackage{dcolumn}
\usepackage{bm}

\usepackage[utf8]{inputenc}
\usepackage[T1]{fontenc}
\usepackage{mathptmx}
\usepackage{amsmath, amsfonts, amssymb, graphicx}
\usepackage{textcomp}
\usepackage{gensymb}
\usepackage{caption}
\usepackage{graphicx, subfig}
\usepackage[mathlines]{lineno}
\usepackage[colorlinks, linkcolor=blue, anchorcolor=blue, citecolor=cyan, urlcolor=cyan]{hyperref}

\usepackage{color}
\usepackage{color, soul}
\soulregister\cite7
\soulregister\citep7
\soulregister\ref7

\usepackage{ulem}

\begin{document}


\title{Deep-learning interatomic potential for irradiation damage simulations in MoS$_2$ with $ab$ $initial$ accuracy}
\thanks{Hao Wang and Xun Guo contributed equally to this work.}

\author{Hao Wang}
 \affiliation{State Key Laboratory of Nuclear Physics and Technology, School of Physics, CAPT, HEDPS, and IFSA Collaborative Innovation Center of MoE College of Engineering, Peking University, Beijing 100871, P. R. China;}

\author{Xun Guo}
 \affiliation{State Key Laboratory of Nuclear Physics and Technology, School of Physics, CAPT, HEDPS, and IFSA Collaborative Innovation Center of MoE College of Engineering, Peking University, Beijing 100871, P. R. China;}



\author{Jianming Xue}
\email{jmxue@pku.edu.cn}
\affiliation{State Key Laboratory of Nuclear Physics and Technology, School of Physics, CAPT, HEDPS, and IFSA Collaborative Innovation Center of MoE College of Engineering, Peking University, Beijing 100871, P. R. China.}
 
\date{\today}

\begin{abstract}
Potentials that could accurately describe the irradiation damage processes are highly desired to figure out the atomic-level response of various newly-discovered materials under irradiation environments. In this work, we introduce a deep-learning interatomic potential for monolayer MoS$_2$ by combining all-electron calculations, an active-learning sampling method and a hybrid deep-learning model. This potential could not only give an overall good performance on the predictions of near-equilibrium material properties including lattice constants, elastic coefficients, energy stress curves, phonon spectra, defect formation energy and displacement threshold, but also reproduce the $ab$ $initial$ irradiation damage processes with high quality. Further irradiation simulations indicate that one single high-energy ion could generate a large nanopore with a diameter of more than 2 nm, or a series of multiple nanopores, which is qualitatively verified by the subsequent 500 keV Au$^+$ ion irradiation experiments. This work provides a promising and feasible approach to simulate irradiation effects in enormous newly-discovered materials with unprecedented accuracy.

\end{abstract}

\maketitle

\section{Introduction}
Since two-dimensional (2D) graphene was fabricated in 2004, investigations of 2D materials were initialized and rapidly developed in recent decades.\cite{RN609,RN579,RN111,RN403,RN110} Among them, monolayer MoS$_2$ has captured tremendous attentions in a very borad area, including nano-electronics,\cite{RN99,RN100} optical sensors,\cite{RN257} catalysts\cite{RN102,RN103} and energy storage.\cite{RN106,RN112} Besides that, ion irradiations/implantations are often applied intentionally to monolayer MoS$_2$ to modulate electrical properties,\cite{RN85,RN376,RN334,RN337,RN352} introduce high-efficiency single-atom catalytic sites\cite{RN336,RN347,RN373} or generate size-controllable nanopores\cite{RN589,RN608,RN365} and so on.\cite{RN85,RN74} However, the mechanism of defect formation in 2D materials under irradiation environment is still controversial, which significantly limits the application of the ion beam technology in modulating 2D material properties. Therefore, understanding the interactions between energetic ions and 2D materials, especially figuring out the formation conditions of the radiation-induced micro-structural defects, is of great importance for extensive potential applications.\cite{RN79}

Nowadays, molecular dynamics (MD) simulation has become a powerful tool for understanding the atomic-level processes which are so small or fast that beyond the limitation of experimental observations, such as the formation and morphology of radiation-induced defects, especially the generation of various nanopores.\cite{RN358,RN366,RN73} However, the validity of MD simulations completely relies on the accuracy of the employed interatomic potential, which is typically a relatively simple analytical function fitted to reproduce a selected set of material properties. \cite{RN465,RN406} However, constructing an interatomic potential that could accurately describe the irradiation processes is a particularly challenging task. A series of atomic-level changes happened during the incident and the following evolution process, including many-body atom collisions, localized melting, rapid recrystallization with extreme temperature and pressure gradients, and defect migrating/clustering and so on.\cite{RN465,RN406,RN390,RN391,RN392,RN453,RN455} Accurately describing all these phenomena apparently beyond the scope of applications for most of the traditional analytical potential models, which are generally designed to describe the near-equilibrium properties of specific systems, e.g., metallic, covalent, or ionic.\cite{RN406} Different potentials could give nearly contradictory simulation results, and sometimes simulated systems may even break down due to abnormal numerical energy predictions of some unstable transition structures.\cite{RN60,RN231,RN232} 

In order to explore high quality MD simulations of monolayer MoS$_2$, several potentials have been established,\cite{RN5,RN2,RN7,RN3,RN241,RN10,RN234,RN236} but none of them is specially constructed for the irradiation collision cascade simulations. For example, previous work has indicated that the predicted defect formation energy and displacement threshold are more or less underestimated or overestimated, and sometimes the irradiated monolayer MoS$_2$ samples will even break down abnormally, \cite{RN73} which can be attributed to the limited fitting parameters and improper function forms of the common analytical potential models. That is to say, some unselected properties might be sacrificed or neglected to satisfy the aiming features. Besides that, the simulated irradiation damage results of two-dimensional materials could be compared quantitatively with experimental observations, especially the formation of size-controllable nanopores, which places a greater demand on the accuracy of potentials employed.\cite{RN608}


Recently, machine learning (ML) methods including the deep neural networks (DNN) are utilized to solve the dilemma between the  questionable accuracy of empirical analytical force fields, and the high computational cost caused by the high accuracy $ab$ $initial$ molecular dynamics (AIMD) with density functional theory (DFT).\cite{RN11,RN34,RN50,RN58,RN67,RN81,RN89,RN90,RN91,RN92,RN93,RN94,RN249,RN306,RN397,RN454} 
Many researches have demonstrated that ML-based potential energy surface can reach the accuracy of DFT with the cost comparable to classical empirical potentials.\cite{RN56,RN87,RN21,RN22,RN228,RN229,RN354} Considering ML-potentials do not assume a specific interaction function for the concerned system, and their accuracy only depend on the  completeness of training database, they are perfect to describe the evolution of highly distorted systems with complex local micro-structures, such as the phase transitions\cite{RN602,RN607} and the irradiation cascade process.\cite{RN406,RN229}


Based on this thought, we have proposed the ZBL modified deep learning scheme (DP-ZBL) for solid materials in 2019,\cite{RN229} which interpolates smoothly the Ziegler-Biersack-Littmark (ZBL) screened nuclear repulsion potential \cite{RN235} into a newly-developed deep-learning force field.\cite{RN22} The resulting DP-ZBL model can not only provide overall good performance on the predictions of near-equilibrium material properties but also accurately reproduce the strong repulsion at the short distance. However, since the universal repulsive ZBL potential is obtained by performing an averaging fit to the results of Thomas-Fermi quantum mechanical calculations for a large number of systems, it is not accurate enough and should be improved for specific ion-ion interactions.\cite{RN226, RN225} Recent researches indicate that the interatomic potentials from all-electron calculations could give an more excellent agreement with the recent ion-solid scattering experiments, \cite{RN225} so it is reasonable to expect that this scheme can also be applied in the simulation of 2D MoS$_2$ systems.


Therefore in this work, we have introduced the all-electron calculated repulsive potential and the active-learning scheme\cite{RN21} into the origin DP-ZBL model and the accuracy of this repulsive table modified deep-learning interatomic potential (DP-Tab) has been strongly enhanced. Then we applied the improved scheme to construct the DP-Tab potential for the monolayer MoS$_2$ to validate the feasibility and reliability of this method and found better performance on the predictions of lattice constants, elastic coefficients, energy stress curves, phonon spectra, defect formation energy and displacement threshold, compared with other existing atomic potentials of MoS$_2$. Moreover, this newly constructed MoS$_2$ DP-Tab potential could not only reproduce the AIMD irradiation damage processes, but also prove that large-scale nanopores or a series of nanopores could be introduced directly by the single-ion irradiation. These novel phenomena are then verified by our subsequent 500 keV Au$^+$ ion irradiation experiments, which demonstrates the DP-Tab potential model could play a guiding role in the understanding of atomic irradiation processes of enormous newly-discovered materials.

\section{Potential scheme}

  In the origin DP-ZBL model, the universal ZBL repulsive potential is tabled and interpolated into the deep-learning potential in practice.\cite{RN229} Actually it's feasible to choose other repulsive potentials, such as the Moliere potential, Jensen potential and user-fitted repulsive potentials.\cite{RN225} In many previous researches, the universal ZBL potential is hybrid with a near-equilibrium potential for irradiation damage simulations.\cite{RN73,RN358,RN366,RN376} However, recent studies have indicated the choose of ZBL potential cutoff will strongly affect the defect productions.\cite{RN231,RN232} Besides that, more accurate scattering cross-section measurements reveal that the traditional ZBL potential need to be improved.\cite{RN225} However, the traditional DFT calculation based on the frozen-core approximation and pseudopotential scheme is no longer accurate at the short distance.  Recently, the all-electron DFT approach was used to calculate the short-range repulsive interatomic potential and the results showed excellent agreement with the ion-solid scattering experiments.\cite{RN225} Therefore, in this paper, we employ the all-electron calculations based on the DFT-Dmol\cite{RN499,RN500} software to generate the precise short-range repulsive table potential. In order to avoid the disconnection of the energy obtained from the generated repulsive table potential and that learned from the DFT-Vasp\cite{RN480,RN481,RN482,RN483} training dataset, the DFT-Dmol energy is smoothly switch to the DFT-Vasp energy as explained below:

\begin{equation}
 E^{\textrm{Tab}}_i= w_j E^{\textrm{DFT-Dmol}}_i + (1 - w_j) E^{\textrm{DFT-Vasp}}_i, 
\end{equation}
The switch function $w_j$ defined as
\begin{equation}
 w_j = \left\{
 \begin{aligned}
  &1 && r_{ij} < R_c, \\
  &-6 u_j^5 + 15 u_j^4 - 10 u_j^3 + 1 && R_c\leq r_{ij} < R_a, \\
  &0 && r_{ij} \geq R_a,
 \end{aligned}
 \right.
\end{equation}
with the $u_j=\dfrac{r_{ij}-R_c}{R_a-R_c}$. The $r_{ij}$ represents the pair distance from the atom $i$ and the atom $j$. In this way, we avoid the energy and force disconnection between the pair table potential and the DFT-Vasp training dataset, which may occur in the origin DP-ZBL model. Note that it's also feasible to fit a new repulsive potential and make it smoothly connect with the training database as the newly developed Gaussian approximation potential done.\cite{RN228}

\captionsetup[figure]{labelfont=bf}
\captionsetup[table]{labelfont=bf}

\begin{figure}[htbp]
	\includegraphics[width=0.9\textwidth]{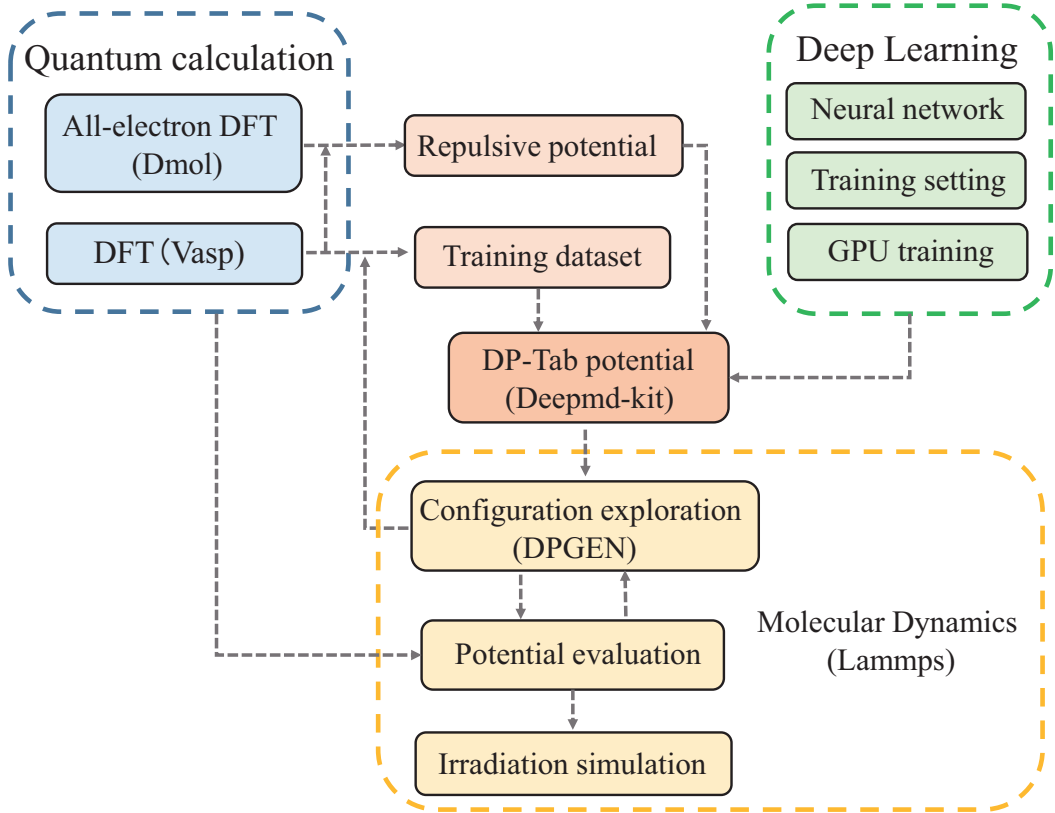}
	\caption{\label{fig:1} The schematic plot of the construction process of the DP-Tab potential.}
\end{figure}

As the schematic diagram shown in FIG.\ref{fig:1}, we have elaborated how to construct the DP-Tab potential. First, the all-electron DFT calculations are used to generate the precise short-range pair repulsive potential, which will be tabled as an input for the DP-Tab potential training. The DFT calculations based on frozen core model and pseudopotential scheme are carried on some representative configurations to generate the initial training dataset. After choosing proper training settings, these inputs are sent to the deep learning neural network\cite{RN22} to generate several rough DP-Tab potentials with different initial seeds. Then the active-learning scheme\cite{RN21} is carried out to run a short-time MD with these potentials at several different temperatures. During this period, the force error indicator is employed to select configurations that could not be consistently predicted, which would be sent to the DFT calculator to get an accurate energy and atomic force. After that, these data will be appended into training datasets and used to retrain the DP-tab potentials. By continually repeating these steps, enormous inaccurately predicted configurations are efficiently picked out and finally contribute to an uniform accurate DP-Tab potential. Then we could use the DFT calculations to evaluate the predications of the DP-Tab potentials on concerned material properties, such as lattice constants, elastic coefficients, equation of states, phonon spectra, defect formation energy,  displacement threshold energy and so on. After verifying that the DP-Tab potential is well trained, we could use it to carry out the corresponding molecular dynamics simulations, especially the irradiation damage events, which is also the focused issue the DP-Tab potential aims to solve. 

Above all, the all-electron DFT calculations and traditional DFT calculations are carried out using the DMol (1997)\cite{RN499,RN500} and VASP (5.4.4)\cite{RN480,RN481,RN482,RN483} software, respectively. The active learning scheme is actually executed with the assistant of the DPGEN (0.7.1)\cite{RN396} software.  The DP-Tab potential model has already been implanted into the Deepmd-kit (1.1.1)\cite{RN354} software, which enables readers to simply develop their own DP-Tab potentials for specific applications.  After the construction of a proper DP-Tab potential, all the molecular dynamic simulations are performed using the LAMMPS\cite{RN489} software. The atomic visualizations are created using the Open Visualization Tool (OVITO)\cite{RN636}. For phonon spectrum calculations, we used the PHONOPY\cite{RN490} code.

\section{Training detail}


\subsection{The DFT-VASP settings}
    All the DFT training structures were calculated using the VASP software and the PBE GGA exchange-correlation functional.\cite{RN484} The Mo 14 $4s^24p^64d^15s^1$ and S 6 $3s^23p^4$ electrons were treated as valence electrons with the core electrons accounted for by the projector-augmented wave (PAW) method\cite{RN485,RN486} (the $\mathtt{Mo\_sv}$ and $\mathtt{S}$ PAW potential in VASP 5.4.4). The plane-wave cutoff energy was 500 eV and the Brillouin zone was integrated using Gamma center grids with a consistent spacing density for all cell sizes (using $\mathtt{KSPACING=0.15}$ \AA$^{-1}$ in VASP). A smearing of 0.1 eV by the first-order Methfessel-Paxton method\cite{RN488} and the $\mathtt{Accurate}$ precision mode were applied to help the convergence. The same settings were used for all our DFT-Vasp calculations in this work. 

\subsection{The repulsive table potential}
First, the DFT-Dmol all-electron method\cite{RN501,RN502,RN503} is utilized to calculate the Mo-Mo, Mo-S, and S-S dimer potential at short-distance 0 $\sim$ 1.2 \AA \ with 0.001 \AA \ step. Although the all-electron calculations fail for dimers whose distance is less than 0.02 \AA \ , such short distances will never be reached even in high-energy cascade simulations, since the pair potential contributes with energies in the MeV range. Second, the DFT-Vasp approach is also employed to calculate the Mo-Mo, Mo-S, and S-S dimer potential at distance among 1.0 $\sim$ 1.2 \AA \  with 0.001 \AA \ step. Then the switch scheme proposed above will be adopted to ensure a smooth connection between the repulsive table potential and the potential learned from the training dataset.

\subsection{The initial training dataset}
\begin{table}
	\caption{\label{table1} Structures included in the initial training database. $N_{\rm s}$ is the number of each structure type, $N_{\rm atoms}$ is the number of atoms in each structure, $N_{\rm atoms}^{\rm tot}$ is the total number of atoms of a given structure type.}
	\begin{ruledtabular}
		\begin{tabular}{cccc}
			Structure type & $N_{\rm s}$ & $N_{\rm atoms}$ & $N_{\rm atoms}^{\rm tot}$  \\
			\hline
			Isolated atom (Mo, S)& 200 & 1  & 200\\
			\hline
			Dimer (Mo-Mo, Mo-S, S-S)& 780 & 2 & 1560 \\
			\hline
			Distorted bulk MoS$_2$ & 7200 & 3 & 21600\\
			\hline
			Bulk liquids &4000 & 54 & 216000\\
			\hline
			Distorted monolayer & 8000 & 3, 12, 27, 48 & 180000 \\
			\hline
			Vacancies ($\rm V_S, V_{Mo}, V_{S^2}, V_{Mo^2}, V_{MoS}, V_{MoS^3}$) & 600 & 104-107 & 63600\\
			\hline
			Substitutional defects ($\rm Mo_S, Mo_{S^2}, S_{Mo}, S^2_{Mo}$) & 400 & 107-109 & 43200\\
			\hline
			Short-range interstitial  & 16000, 840 & 13, 49 & 249160 \\
			\hline
			Nanopores  & 29700 & 40-47 & 1285200 \\
		\end{tabular} 
	\end{ruledtabular}
\end{table}
After generating the proper repulsive table potential, the second step is to generate the initial training dataset. As shown in the Table.\ref{table1}, we have selected several configurations as the initial selected structures, including isolated atoms, dimers, distorted bulk MoS$_2$ unit cells, bulk liquids structure, distorted monolayer unit cells, vacancies defects, substitutional defects, short-range interstitials, and nanopore defects. Below we will explain the reasons why we choose these structures as initial training datasets. 

Since our purpose is to develop an accurate deep-learning interatomic potential to study the irradiation damage processes of the monolayer MoS$_2$, the pristine and distorted monolayer unit cells with compression ratio of 0.5 $\sim$ 1.5  are  essential structures to get the near-equilibrium and high compress/stretch potential energy surface.  Usually, various defects could be formed during the irradiation processes, including the representative vacancy defects ($\rm V_S, V_{Mo}, V_{S^2}, V_{Mo^2}, V_{MoS}, V_{MoS^3}$), substitutional defects ($\rm Mo_S, Mo_{S^2}, S_{Mo}, S^2_{Mo}$), and nanopores which have already been observed experimentally.\cite{RN365,RN349,RN340,RN344} For the vacancy and substitutional defects, we simply include these structures. In order to describe various nanopores formed during the irradiation damage process, we have also added these 4x4 unit cells with Mo$_{n}$S$_{m}$ (n=0, 1, 2; m=0, 1, 2, 3, 4, 5, 6) nanopores. In order to adjust the defect formation energy and reproduce the correct cohesion, the isolated atoms structures are also included. Moreover, local melting and recrystallization should be well described to reproduce atomic mixing during the heat spike of a collision cascade. Therefore, we have prepared the distorted bulk and liquid configurations. We considered a set of random configurations generated using the PACKMOL\cite{RN504} software around the experimental density of liquid bulk MoS$_2$ 4.8 g/cm$^3$. To ensure a physical reasonable dissociation of atoms as well as to smoothly connect to the repulsive table potential, we have also included the Mo-Mo, Mo-S, S-S dimers among 1.2 $\sim$ 5.0 \AA\ with 0.1 \AA\ step. To capture the short-range many-body dynamics in the monolayer MoS$_2$ during the collision cascades, we have prepared enormous structures with a randomly added interstitial Mo/S atom (called "short-range interstitials" in Table.\ref{table1}). The shortest allowed distance between the added atoms and its neighbors was 1.2 \AA, in accordance with the switch range between the repulsive table potential and the potential learned from the training dataset. It is worthy noted that all these initial training datasets are prepared using the $init\_bulk$ module implanted in the DPGEN software by running a short-time AIMD.\cite{RN396} The atoms perturbation and box shift as well as the active-learning processes are automatically done by setting the corresponding key parameters, which strongly reduces manual operations to train a reliable DP-Tab potential.

\subsection{The training process}
After finishing the two necessary inputs: the repulsive table potential and the initial training datasets, we should choose proper training settings to make the potential neural network easily converged. The loss function of the origin DP model\cite{RN22} is defined as
\begin{equation}
L\left(p_{\epsilon}, p_{f}, p_{\xi}\right)=\frac{1}{|B|} \sum_{l \in B} p_{\epsilon}\left|E_{l}-E_{l}^{w}\right|^{2}+p_{f}\left|F_{l}-F_{l}^{w}\right|^{2}+p_{\xi}|| \Xi_{l}-\Xi_{l}^{w} \|^{2}
\end{equation}
Here $B$ denotes the minibatch, $|B|$ is the batch size, $l$ denotes the index of the training data, see ref.[\citet{RN22}] for more detail. Since we have added the short-range dimers and interstitials configurations into our training datasets, whose atomic forces are actually near 100 eV/\AA\ and thus 3$\sim$4 orders of magnitude than those in the equilibrium states, which results in the hard convergence of the training process. Considering that the relative atomic force error is more meaningful than the absolute atomic force error for these short-range structures, we have employed the relative force error in this work. The relative force error on atom $i$ is defined as
\begin{equation}
E(f_i)=\dfrac{|F_i-F_i^w|}{|F_i|+level}
\end{equation}
Here the $F_i$ represents the accurate atomic force in the training dataset, $F_i^w$ denotes the atomic force predicted by the current DP-Tab model and the $level$ is a protection constant and fixed to 1 in this work. In Fig.S1, we have plot the force loss error during the training process using the same training dataset, which indicates the relative force error is much better for the convergence of the training process than the previous adopted absolute force error in this work. 

After choosing proper settings for the fitting neural network, we begin to get into the active-learning schemes. In practice, four DP-Tab potentials with random seeds are trained with the initial training dataset. Then the multiple temperature (0 $\sim$ 1200 K) NVT MD simulations are carried out on several structures obtained from the initial dataset. During this period, the force error indicator is adopted to select the inaccurately predicted structures.\cite{RN21} The accurately predicted structures are refereed to the maximum atomic force deviation from the four trained potential models is below 0.05 eV/\AA.\ If the calculated atomic force deviation is in the range 0.05 $\sim$ 0.15 eV/\AA,\ the structures are selected as candidates and sent to the DFT-Vasp calculator. After the converged DFT calculations, the corresponding energy and atomic force will be extracted and appended into the training dataset. Then the four DP-Tab potential models will be retrained iteratively. We have carried out more than 200 iterations, more than 100000000 structures have been explored and among them more than 45000 candidate structures are selected and expanded into the training dataset after the DFT calculations. Considering enormous configurations included in the training datasets, great attempt is undergoing to minimize the effort to construct an uniform DP-Tab potential. Furthermore, we have summarized all the training parameters used in this work in Table.S2.

\section{Results and Discussions}
\begin{figure}[htbp]
\centering
\begin{minipage}[t]{0.8\textwidth}
\centering
\includegraphics[width=1.0\textwidth]{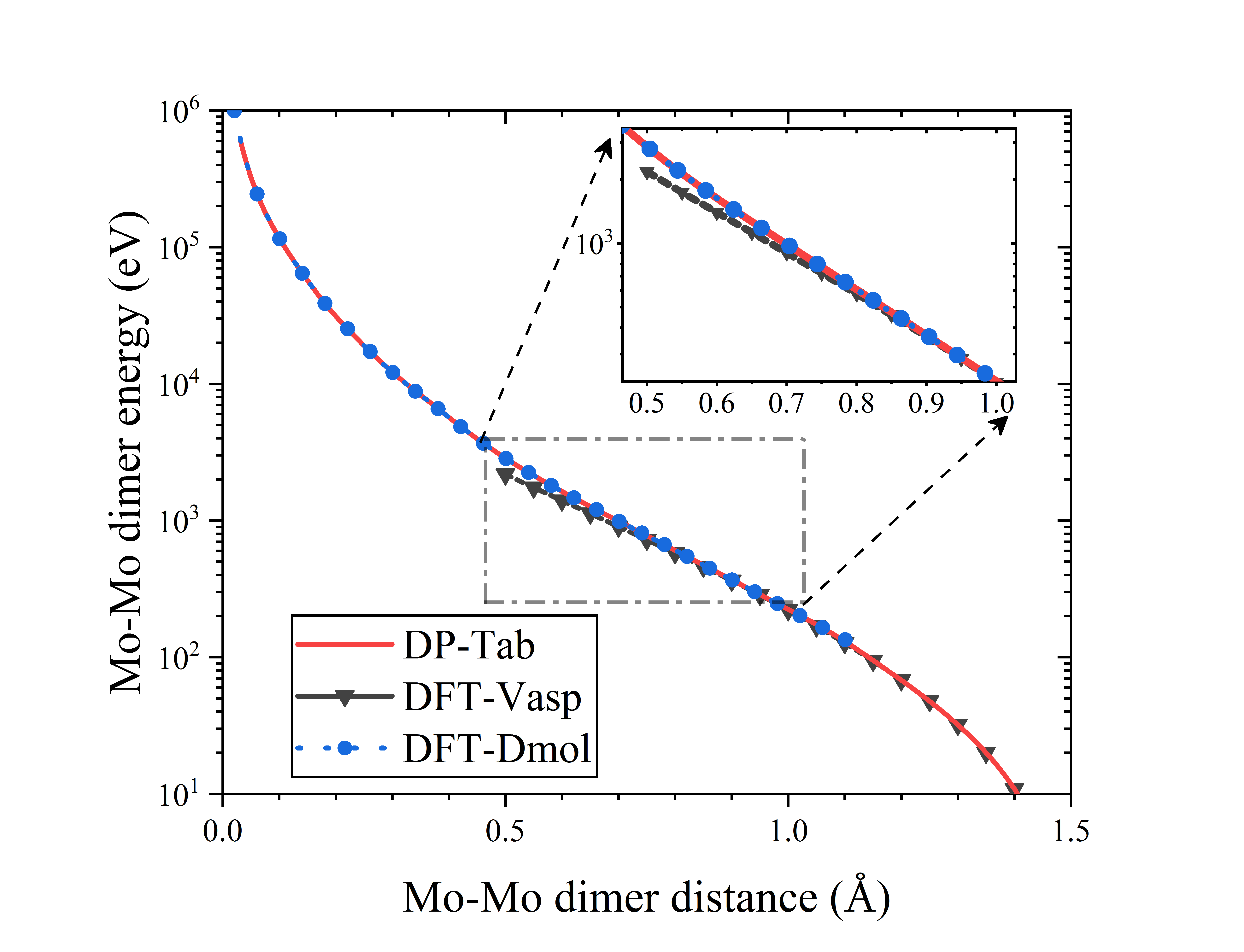}
\end{minipage}%
\quad
\begin{minipage}[t]{0.8\textwidth}
\centering
\includegraphics[width=1.0\textwidth]{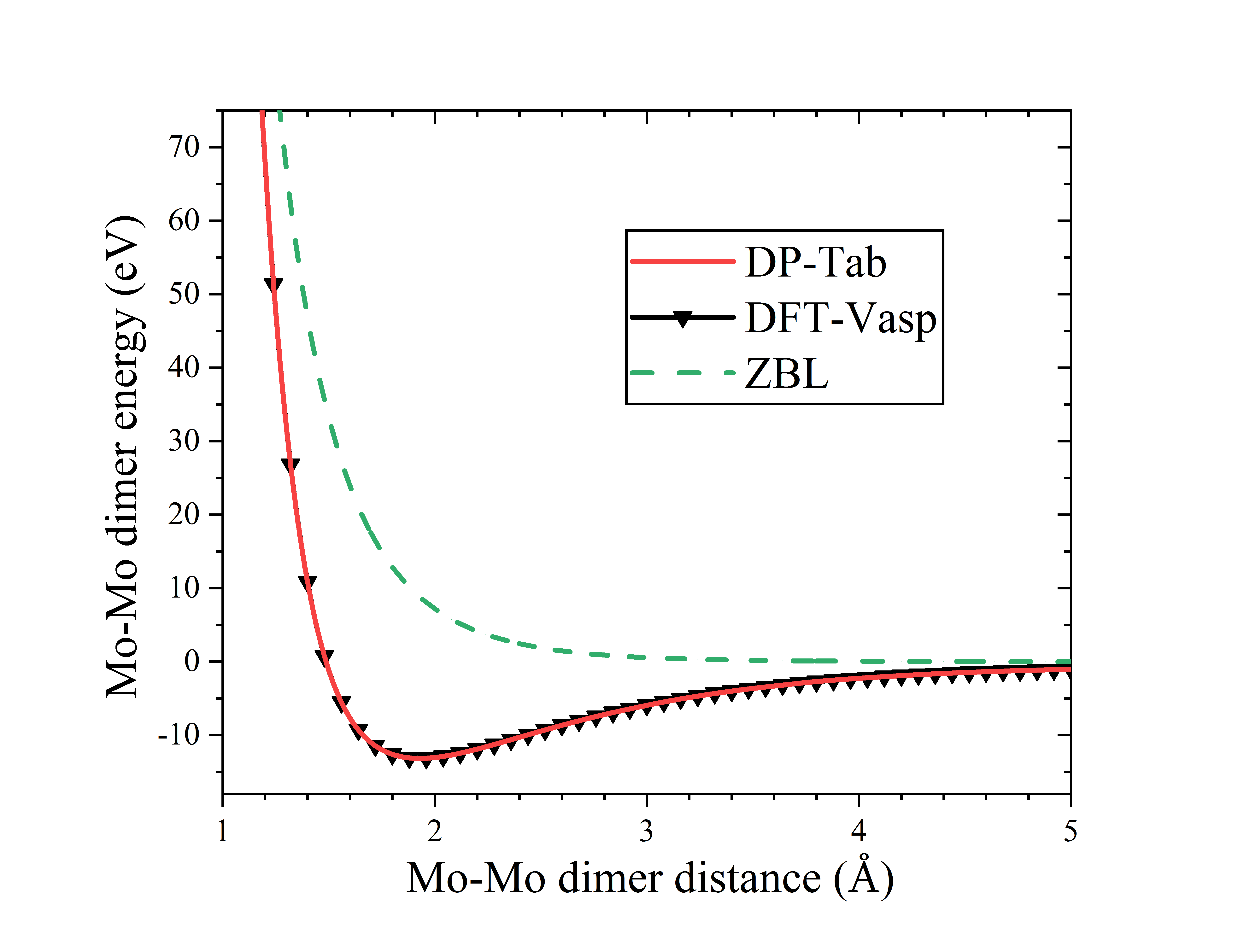}
\end{minipage}%

\centering
\caption{\label{fig:2}The calculated Mo-Mo dimer energy-distance curves based on the all electron DFT method (DFT-Dmol), the pseudopotential DFT method (DFT-Vasp), the generated DP-Tab potential and the universal ZBL repulsive potential.}
\end{figure}
After the active learning scheme, we begin to evaluate the performance of the proposed DP-Tab potential on the concerned material properties. The behaviors of the classical SW-FM potential\cite{RN10} and the machine-learning SNAP potential\cite{RN236} are also assessed for comparison, which both are optimized toward a larget set of DFT training datasets. First of all, we calculate the dimer energy-distance curve to validate that the generated DP-Tab potential could smoothly switch from the all-electron repulsive table potential to the DFT-Vasp obtained potential. As shown in Fig.\ref{fig:2}, the DFT-Vasp result is no longer accurate for  the Mo-Mo dimers less than 1.0 \AA, due to the frozen core approximation and pseudopotential method as compared with the DFT-Dmol all-electron method, which also indicates the choose of (1.0, 1.2) \AA\ as the table potential switch range is reasonable. For the Mo-Mo dimer whose distance is larger than 1.2 \AA, the DP-Tab potential gives a perfect agreement with the DFT-Vasp results, which demonstrates that the dissociation of atoms could be well described.

Then we calculated some near-equilibrium material properties as summarized in Table.\ref{table3}. The defect formation energy is defined as $E_f = E_{\rm def} -E_{\rm bulk} + n\mu_{X}$, where $E_{\rm bulk}$ and $E_{\rm def}$ are the energy of the pristine monolayer system and corresponding defect structures, respectively. The chemical potential $\mu_{X}$ of the Mo/S species is taken as the energy of the isolated atom in this work.\cite{RN311} It's no wonder that all the three potentials provide reliable lattice constants $a_0$, $b_0$ and Mo-S bond length d(Mo-S) as the pristine structures are utilized to fit the tunable parameters. As for the vacancy formation energy
$E_{\rm f}$(V$_{\rm S}$) and $E_{\rm f}$(V$_{\rm Mo}$), both the DP-Tab and SW-FM potentials give accurate predications. However, the SNAP potential overestimates the $E_{\rm f}$(V$_{\rm S}$) and even results in the system breakdown when calculating the $E_{\rm f}$(V$_{\rm Mo}$). By looking through the training dataset of the SNAP potential, we find that the bad performance on defective structures of the SNAP potential is reasonable, since no defective configurations are included in the training datasets. It reminds us that the machine-learning potentials may give a bad or even wrong prediction for structures far from the training datasets and additional evaluation is essential before carrying out the concerned MD simulations.\cite{RN406} We have also calculated the elastic properties of the monolayer MoS$_2$. The in-plane Young Modulus (Y) and Poisson's ratio ($\nu$) are extracted from the following relations : Y = (C$_{11}^2$ -  C$_{12}^2$)/C$_{11}$ and $\nu$ = C$_{12}$/C$_{11}$.\cite{RN325} Both the DP-Tab and SNAP potentials give a reliable description on these elastic properties, while the SW-FM potential underestimates C$_{11}$ and overestimates the C$_{12}$, which results in the low Y and the high $\nu$ compared to the DFT calculations.

\begin{table}
	\caption{\label{table3}The calculated material properties of the monolayer MoS2: equilibrium lattice constants a$_{0}$, b$_{0}$ and the distance between two chalcogen atoms above and below the Mo layer $d$, the Mo-S band length $d(\rm Mo-S)$, vacancy formation energy $E_{\rm f}$(V$_{\rm S}$) and $E_{\rm f}$(V$_{\rm Mo}$), and independent elastic coefficients $C_{\rm 11}$ and $C_{\rm 12}$, young modulus Y, Poisson's ratio $\nu$.}
	\begin{ruledtabular}
		\begin{tabular}{cccccc}
MoS$_2$  & DFT (this work) & DP-Tab & SNAP & SW-FM & DFT(ref) \\
			\hline
a$_{0}$ (\AA)  &3.18 & 3.18 &3.14 & 3.20 & 3.18\cite{RN3}\\                                    
            \hline			
b$_{0}$ (\AA)  &3.18 & 3.18 &3.14 & 3.20 & 3.18\cite{RN3} \\
			\hline
$d$ (\AA)      &3.12 & 3.12 & 3.12 & 3.19 & 3.11\cite{RN234}   \\
			\hline
$d (\rm Mo-S)$  &2.41 & 2.41 & 2.39 & 2.44 & 2.43\cite{RN234}\\
            \hline                  
$E_{\rm f}$ (V$_{\rm S}$) &6.7 & 6.5 & 8.4  & 6.6 & $\sim$ 7.0\cite{RN73} \\
			\hline
$E_{\rm f}$ (V$_{\rm Mo}$)& 17.8 &17.0 & - &17.5 & $\sim$ 19 \cite{RN73}\\
			\hline
$C_{\rm 11}$ (N/M) & 132.12 & 128.03 & 135.24 & 117.71 & 132.7\cite{RN3}\\
			\hline
$C_{\rm 12}$ (N/M) & 31.92  & 27.44  & 34.38  & 40.52  & 33.0\cite{RN3}\\
		    \hline
			Y (N/M) & 124.41 & 122.15 & 126.50 & 103.76 & 124.5\cite{RN3}\\
		    \hline
            $\nu$  & 0.24   & 0.21   & 0.25   & 0.34   & 0.25\cite{RN3}\\
		    
		\end{tabular}
	\end{ruledtabular}
\end{table}

\begin{figure}[htbp]
	\includegraphics[width=0.9\textwidth]{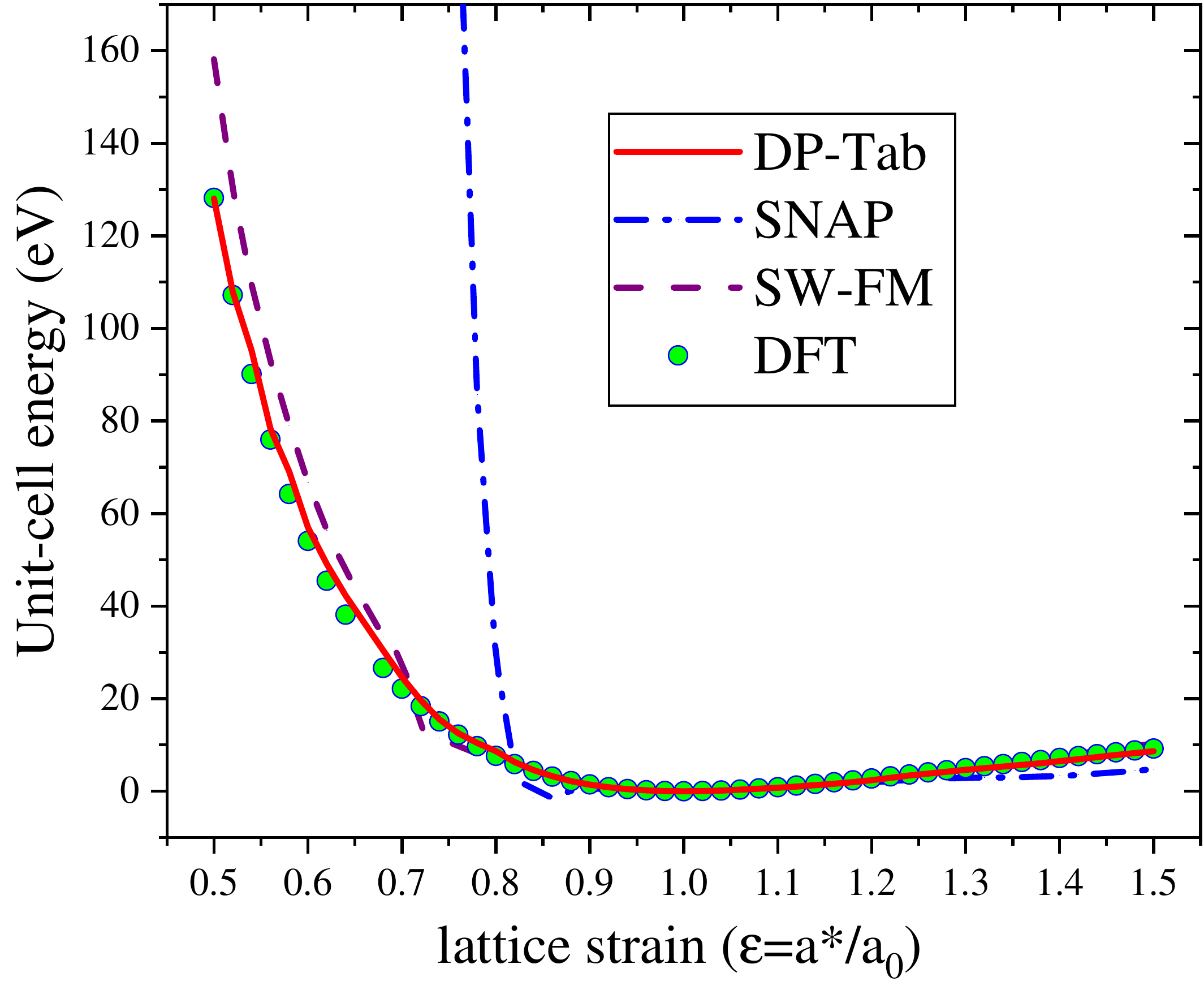}
	\caption{\label{fig:3} The energy-strain curves based on the DP-Tab, SNAP, SW-FM potentials and the DFT calculations. The unit-cell energy of the pristine system is referred as 0.}
\end{figure}

Then we calculate the energy-strain curves of the monolayer MoS$_2$. As shown in Fig.\ref{fig:3}, the DP-Tab potential shows the perfect agreement with the DFT-Vasp results, which indicates the DP-Tab potential is sufficient to describe the high compressing and stretching situations during the irradiation damage events. However, the SW-FM and SNAP potentials overestimate the corresponding strain energy when compressing the lattice. Note that the SNAP potential even gives a negative value at about 0.85 lattice strain, which is obviously against the experimental observations and theoretical calculations.

\begin{figure}[htbp]
\centering
\begin{minipage}[t]{0.5\textwidth}
\centering
\includegraphics[width=1.0\textwidth]{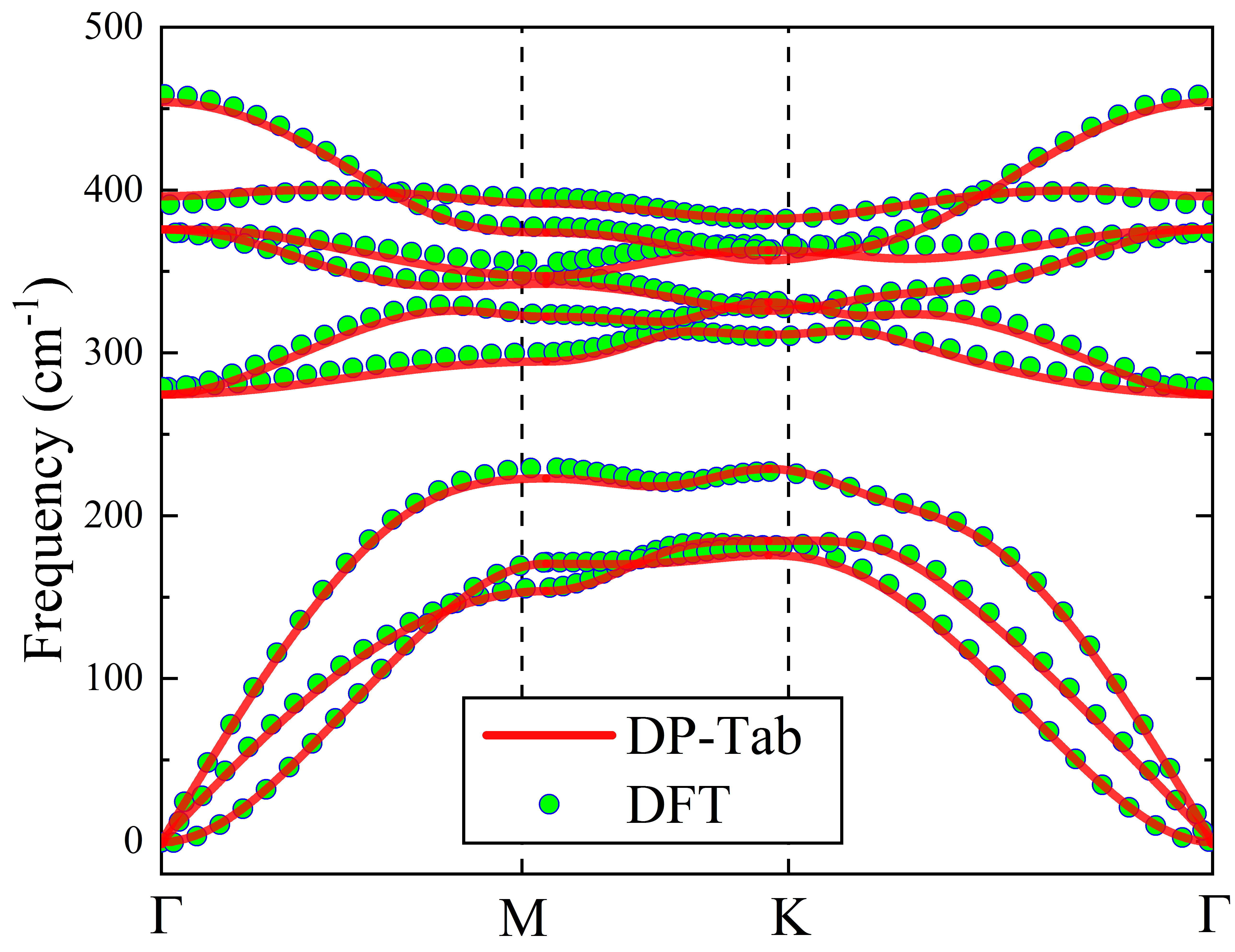}
\end{minipage}%
\begin{minipage}[t]{0.5\textwidth}
\centering
\includegraphics[width=1.0\textwidth]{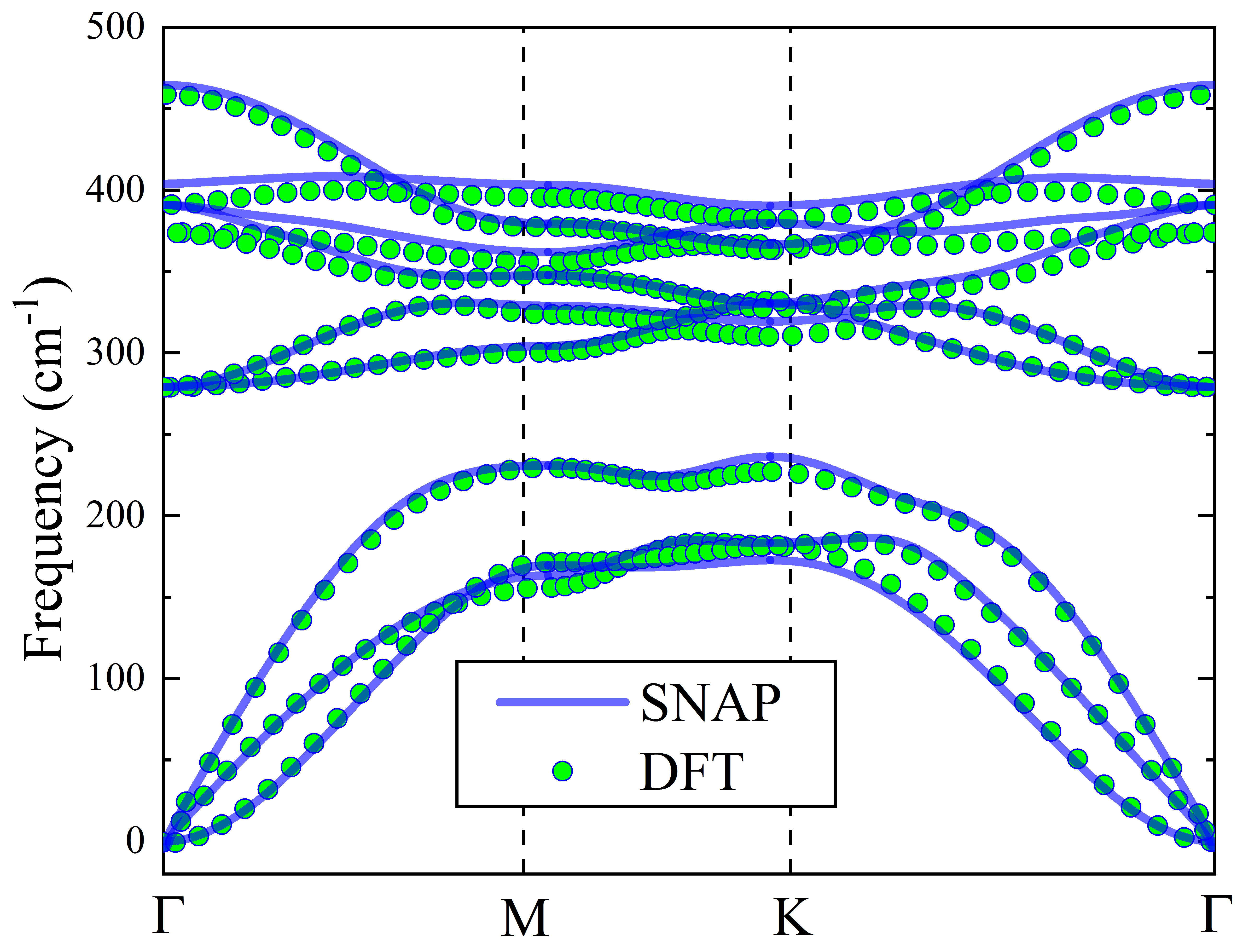}
\end{minipage}%
\quad
\begin{minipage}[t]{0.5\textwidth}
\centering
\includegraphics[width=1.0\textwidth]{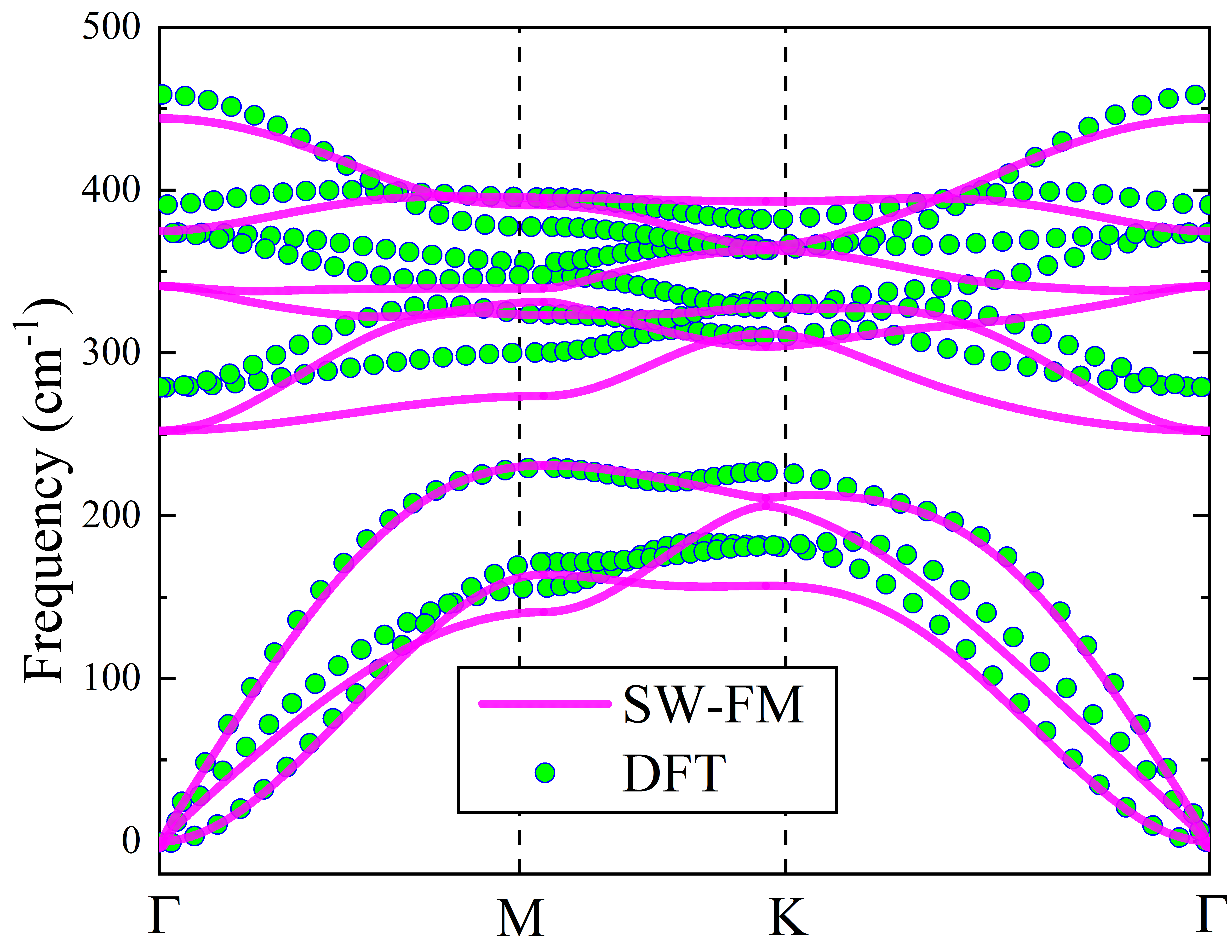}
\end{minipage}%
\centering
\caption{\label{fig:4}The phonon spectra of the monolayer MoS$_2$ based on the DP-Tab, SNAP, SW-FM potentials and the DFT calculations, respectively.}
\end{figure}
Fig.\ref{fig:4} shows the phonon dispersion of monolayer MoS$_2$ obtained using the DP-Tab, SNAP, SW-FM potentials and the DFT method, which could represent the lattice vibration of different modes. The dispersion relation is overall well-reproduced by the DP-Tab and SNAP potential. Note that the SNAP potential is specifically designed for the thermal conductivity of monolayer MoS$_2$, while the DP-Tab potential learns the vibrational properties directly from the training datasets without human intention. The phonon spectrums of the REBO2009,\cite{RN5} SW2013,\cite{RN2} and SW2016\cite{RN3} have been investigated in previous studies.\cite{RN1,RN236} However, these potentials could only give a rough description of the phonon dispersion as well as the SW-FM potential adopted in this work, which demonstrates that the limited tunable fitting parameters of classical analytical potentials may restrict the ability to reproduce the enormous lattice vibration modes. The above calculations indicate that the proposed DP-Tab potential could give an overall good performance on the material properties near the equilibrium and reproduce the excellent repulsive potential at the short distance.
\begin{figure}[htbp]
	\includegraphics[width=0.9\textwidth]{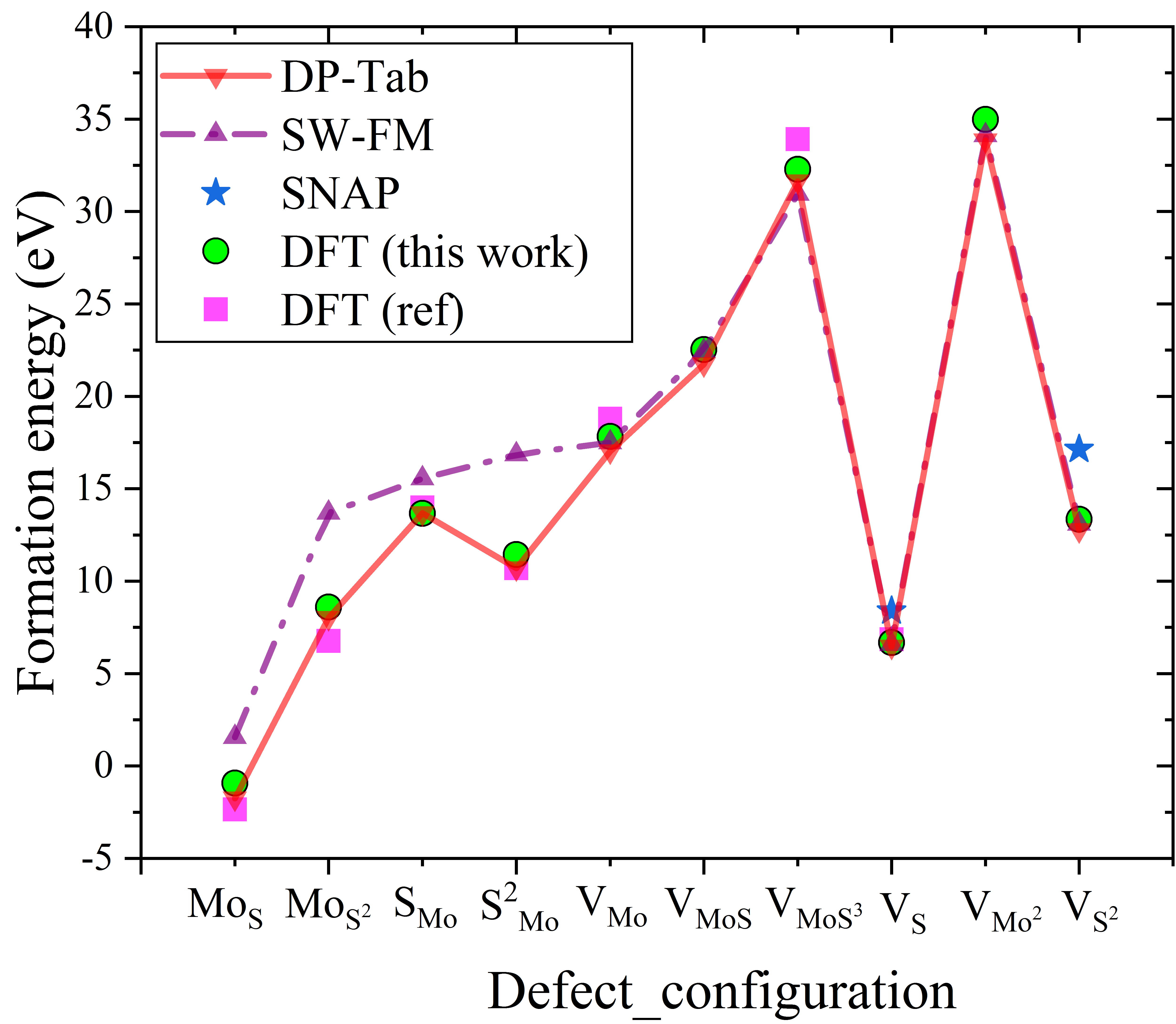}
	\caption{\label{fig:5} The calculated defect formation energy based on the DP-Tab, SNAP, SW-FW potentials and the DFT calculations, respectively.}
\end{figure}

Before employed into irradiation damage simulations, the good performance on the various defect formation energy is required. Getting the defect energies correct is key to reliably simulating the primary damage states produced by energetic recoils.\cite{RN406}  As shown in Fig.\ref{fig:5}, the DP-Tab potential gives an excellent agreement with the DFT results for all the vacancy and substitutional defects considered in this work. The SW-FM potential gives a good performance on the vacancy type defects, including V$_{\rm Mo}$, V$_{\rm MoS}$, V$_{\rm MoS^3}$, V$_{\rm S}$, V$_{\rm Mo^2}$, V$_{\rm S^2}$, while overestimates the formation energy for substitutional defects, including Mo$_{\rm S}$, Mo$_{\rm S^2}$, S$_{\rm Mo}$, S$^2_{\rm Mo}$. As for the SNAP potential, we only work out the formation energy for V$_{\rm S}$ and V$_{\rm S^2}$. Unfortunately, for the structures with other defects, the SNAP potential gives an unphysical low energy predication, which results in the abnormal formation energy over thousands eV. Since these defects are not included in the training dataset of the SNAP, it's not surprised to see the SNAP potential gives a terrible description of the defective configurations. Therefore, we should examine the performance of the constructed potential on the concerned material properties before carrying out the corresponding MD simulations, in order to avoid the inveracious results. Previous researches have also calculated the formation energy of V$_{\rm S}$, V$_{\rm Mo}$, V$_{\rm MoS_3}$, S$_{\rm Mo}$, Mo$_{\rm S}$, Mo$_{\rm S^2}$, S$^2_{\rm Mo}$ using the REBO2009, SW2013, SW2015, SW2016 potentials,\cite{RN73} while none of these potentials behaves better than the DP-Tab potential proposed in this work.
\begin{table}
	\caption{\label{table4} The average displacement threshold energy $E_{\rm d}$ (eV) of Mo, S atoms in the monolayer MoS$_2$.}
	\begin{ruledtabular}
		\begin{tabular}{ccccccc}
			Method & DFT & DP-Tab & SNAP & SW-FM &SW2013 & REBO2009 \\
			\hline
			$E_{\rm d}$ (Mo) & 20\cite{RN311} & 19  & - & 34 & 31.7\cite{RN73} & 38.9\cite{RN73}\\
			\hline
			$E_{\rm d}$ (S)  & 6.9\cite{RN311} & 7.0 & - & 7.5 & 5.0\cite{RN73} & 9.3\cite{RN73} 
		\end{tabular}
	\end{ruledtabular}
\end{table}

Then we calculated the displacement threshold energy E$_d$ of the monolayer MoS$_2$, which is often the key parameter to evaluate the defect production during the collision cascades. As shown in TABLE.\ref{table4}, the DFT calculated E$_d$ is about 20 eV for Mo and 6.9 eV for S.\cite{RN311} The DP-Tab potential gives the best description among the concerned potentials . The SW-FM potential gives a reliable result for S threshold energy, but overestimates the E$_d$(Mo). As for the SNAP potential, the systems generally break down before we could access the displacement threshold energy, due to the terrible energy prediction of defective structures.

\begin{figure}[htbp]
\centering
\begin{minipage}[t]{0.5\textwidth}
\centering
\includegraphics[width=1.0\textwidth]{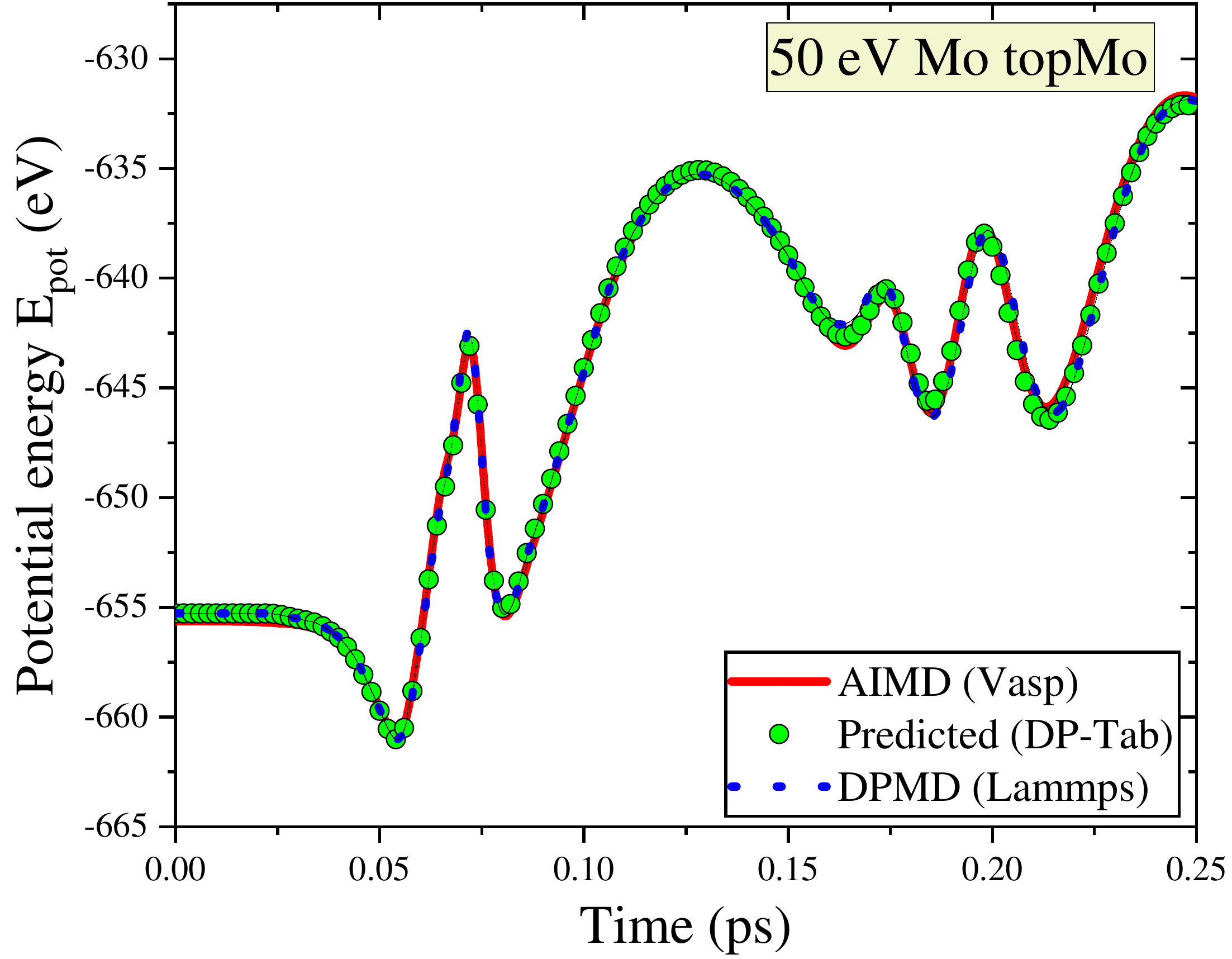}
\end{minipage}%
\begin{minipage}[t]{0.5\textwidth}
\centering
\includegraphics[width=1.0\textwidth]{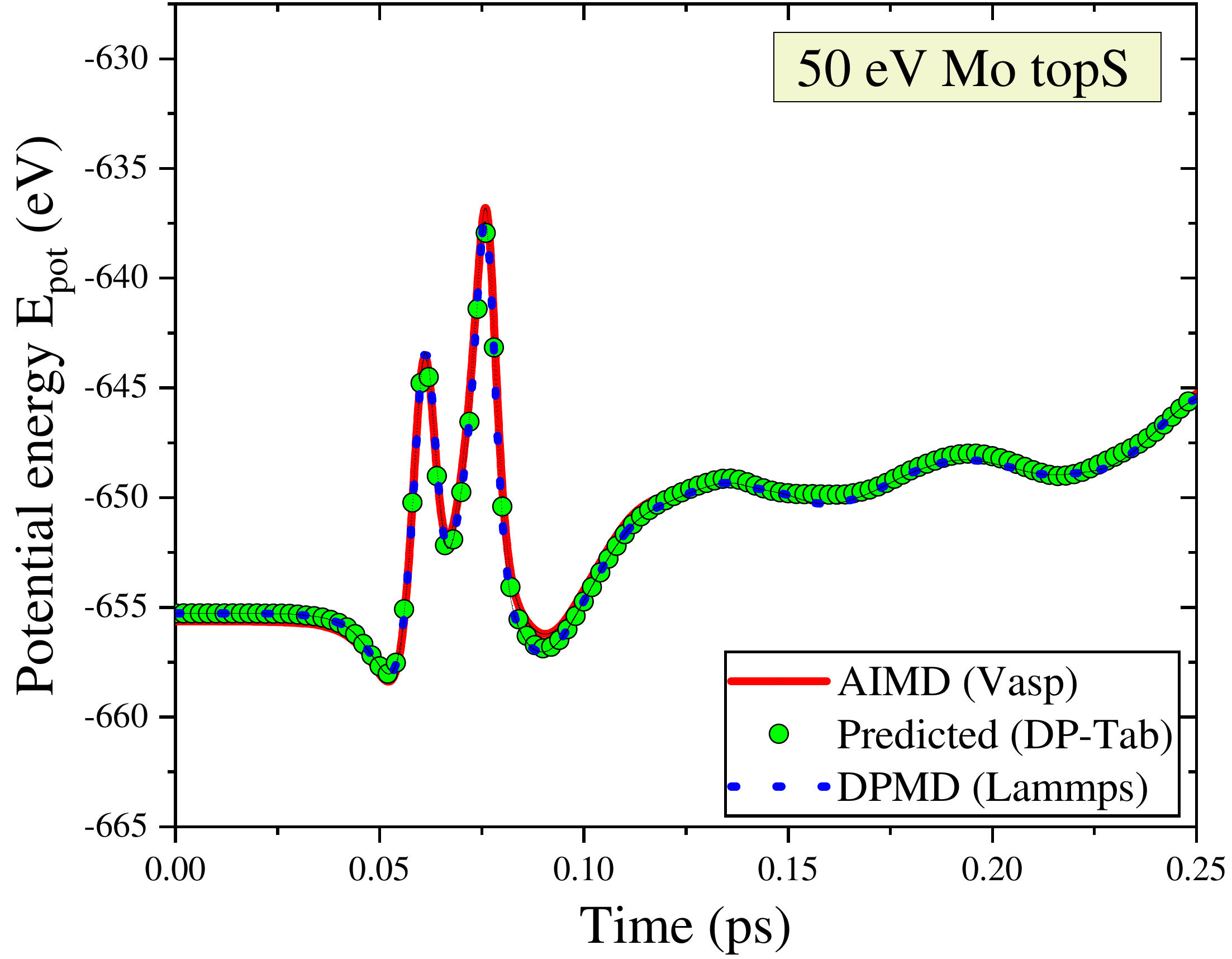}
\end{minipage}
\quad
\begin{minipage}[t]{0.5\textwidth}
\centering
\includegraphics[width=1.0\textwidth]{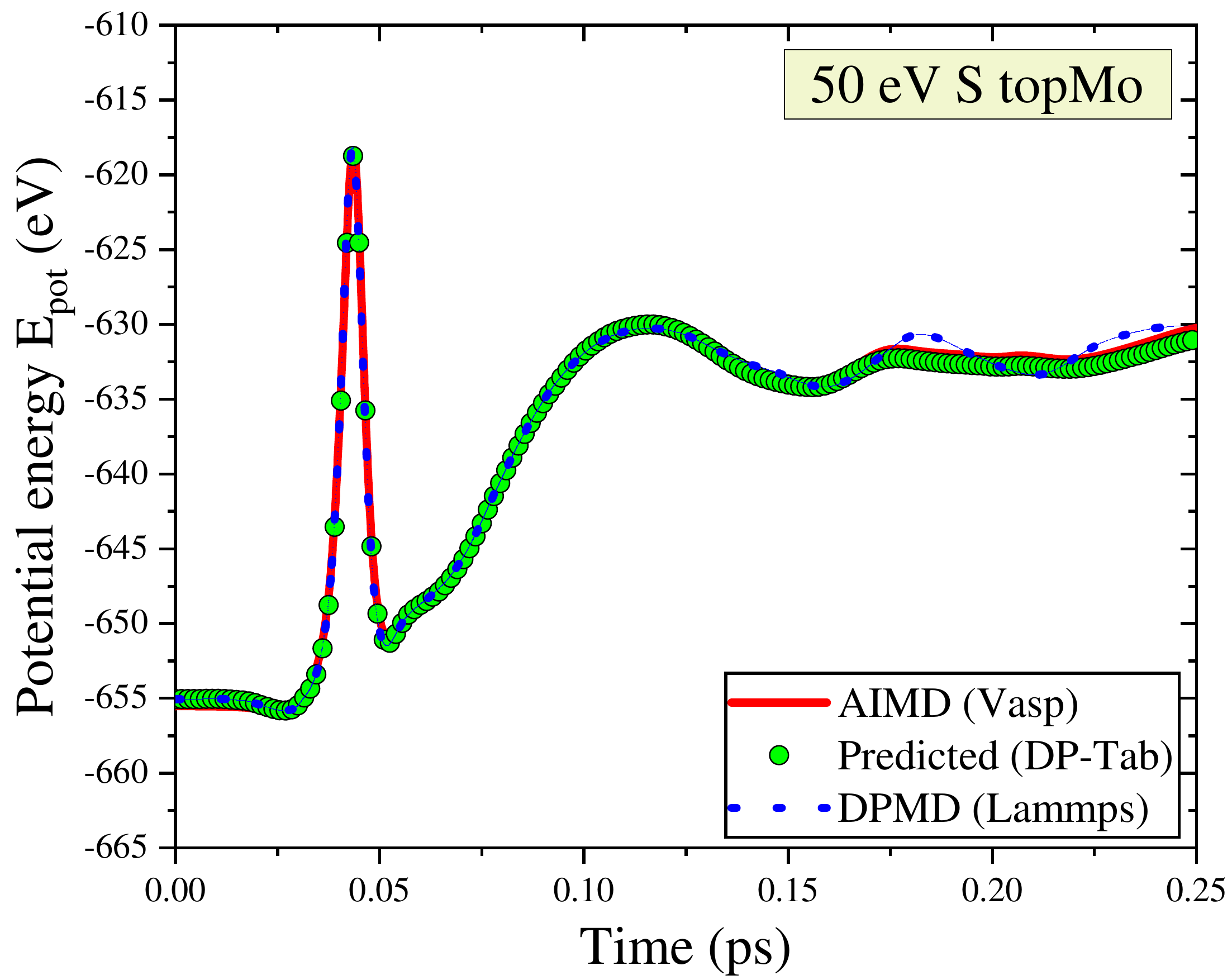}
\end{minipage}%
\begin{minipage}[t]{0.5\textwidth}
\centering
\includegraphics[width=1.0\textwidth]{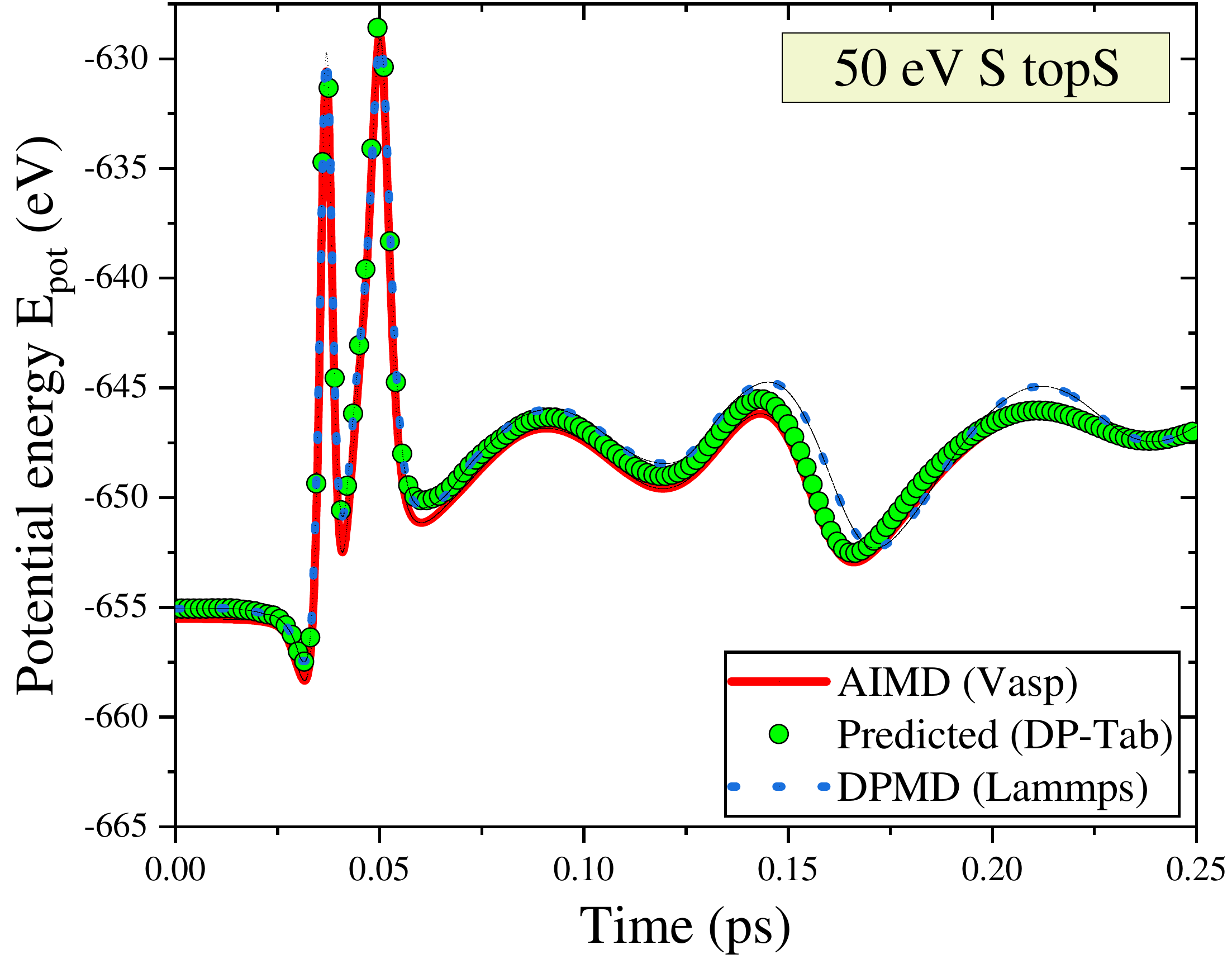}
\end{minipage}

\centering
\caption{\label{fig:6}The potential energy change during the bombardment of 50 eV (a) Mo and (b) S PKAs into the free $\rm{MoS_2}$ monolayer. TopMo means the incident atom is right above the target Mo atom.}
\end{figure}

In order to validate that the generated $\rm{MoS_2}$ DP-Tab potential could describe the irradiation damage processes with the $ab$ $initial$ accuracy, we have simulated the bombardment processes of 50 eV Mo/S primary knock-on atoms (PKAs) into free $\rm{MoS_2}$ monolayers. As shown in the Fig.\ref{fig:6}, the AIMD simulations are in good agreement with classical MD simulations using the DP-Tab potential by visualizing the damage events and monitoring the potential energy of the whole system. Besides that the DP-tab potential could give nearly the same energy predictions of the generated structures during the irradiation damage processes from the AIMD trajectories, which indicates that the well-established DP-Tab potential could reproduce the irradiation damage processes with $ab$ $initial$ accuracy. Note in this work, we neglected the excitation and ionization effect of the electrons during the high-energy cascade, due to the classical DFT calculation is not sufficient for these problems while the time dependent density function theory (TDDFT) calculations are still too expensive now.

Finally, we have also performed extensive damage simulations of Mo/S ion irradiation events. For each energy point, 100 independent simulations are carried out to get average results. The incident point is located randomly within a unit cell of the lattice. \cite{RN73} (For more detail about the simulation systems, please refer to the Supplementary Materials)
As shown in Fig.7, the total sputtering yields firstly increase and then decrease with the incident energy as confirmed in many previous studies, due to the synergistic effect of ion energy and collision cross section.\cite{RN73,RN366,RN358} Besides that, in most energy level considered, the damage produced by the incident Mo ions is larger then S ions, because of its higher mass and rather longer interaction time. The sputtered S is much more than the sputtered Mo, considering its rather lower E$_d$ and higher proportion, which has been already observed in many previous irradiation experiments and simulations.\cite{RN73,RN589,RN74,RN347} Considering the initial direction of the incident ions, the sputtered S from the bottom layer is usually far more than the bottom layer. However, it is worthy noted that, at the 10eV and 20 eV, the sputtering yield of S incidents is more than the Mo incidents. That is because at low energy, the initial ion is hardly to get through the monolayer, so the sputtering yield is dominated by the top layer S. According to the binary collision approximation (BCA) theory, the S ions could transfer more energy to the target S atoms than the Mo ions. 
\begin{figure}[htbp]
	\includegraphics[width=0.9\textwidth]{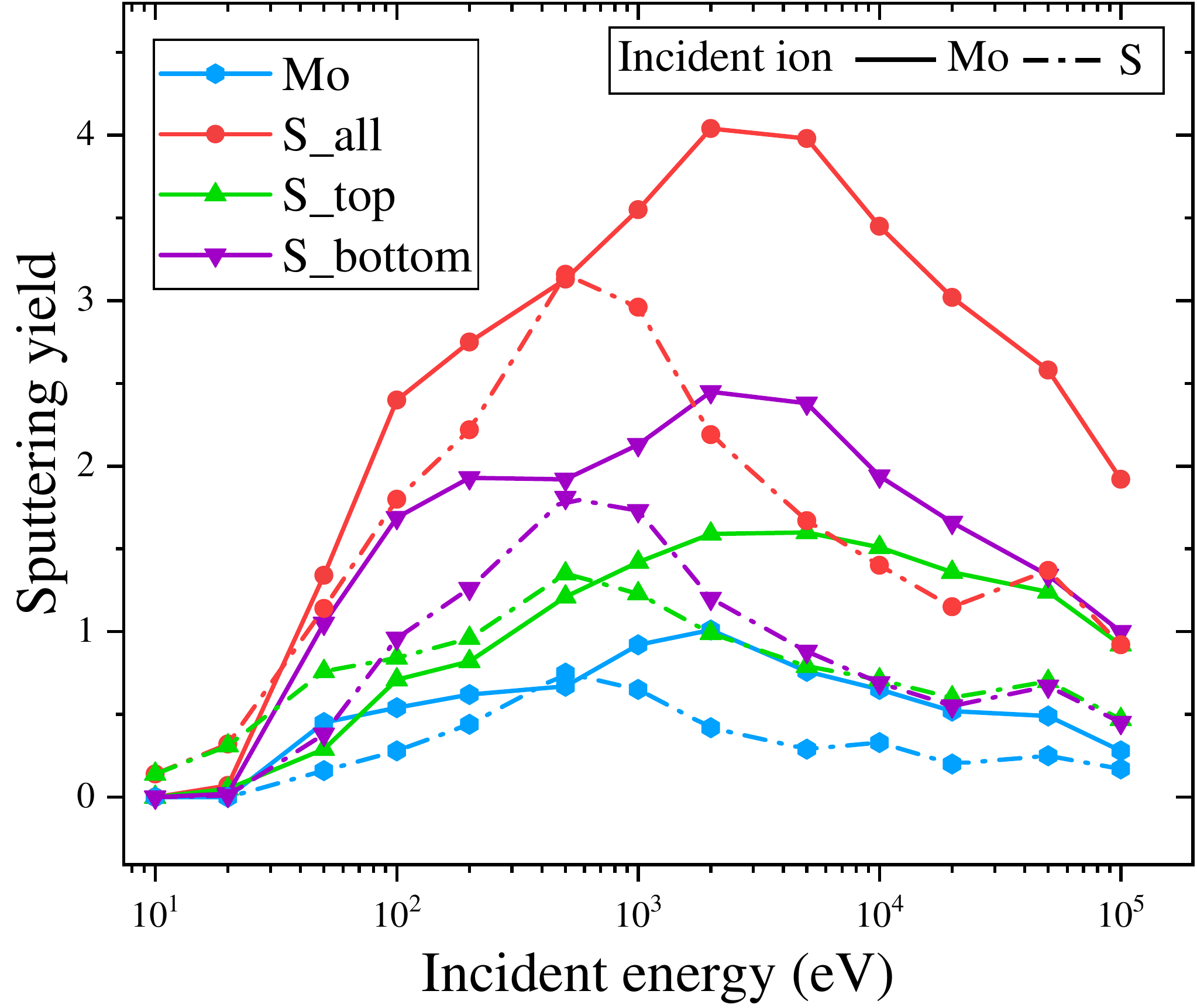}
	\caption{\label{fig:7} The simulated sputtering yield as a function of the incident energy. Each point is average of 100 independent simulations.}
\end{figure}

\begin{figure}[htbp]
	\includegraphics[width=1.0\textwidth]{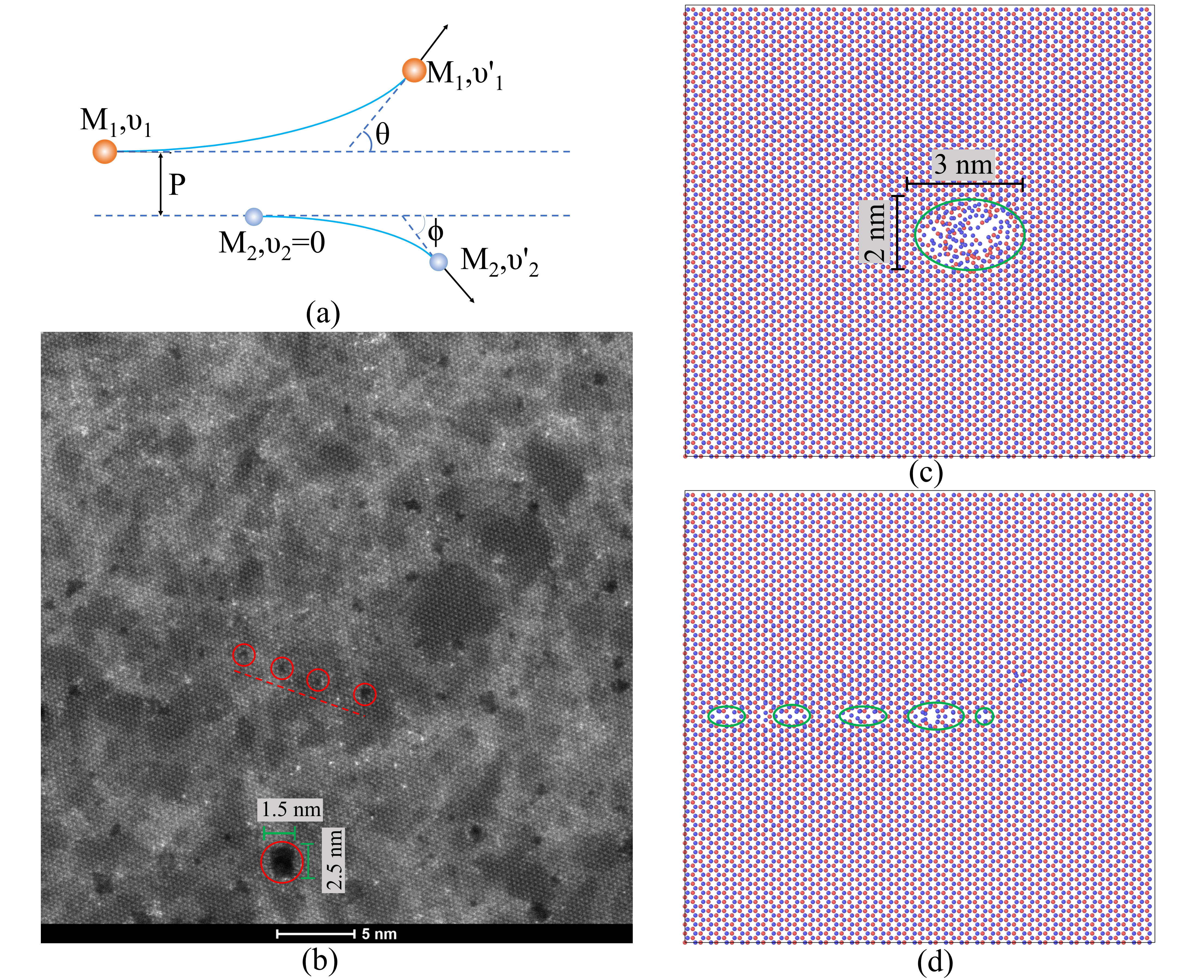}
	\caption{\label{fig:8} (a) The schematic plot of a binary collision. (b) The free monolayer MoS$_2$ after a dose 3$\times$10$^{13}$ of 500 keV Au$^+$ irradiation. (c) and (d) Simulated nanopores in monolayer MoS$_2$ after 500 keV Au$^+$ irradiation.}
\end{figure}

Although the average sputtering yield of high energy incidents is rather low, according to our 500 keV Au$^+$ simulation results, sometimes rather a large nanopore with a diameter of more than 2 nm or a series of multiple nanopores could be produced directly when one single high-energy ion induces interlayer collision cascades as shown in Fig.\ref{fig:8}(c) and (d). These novel phenomena have never been observed and reported before. Therefore, we explore this problem by using 500 keV Au$^+$ ions to irradiate the freestanding monolayer MoS$_2$  samples without the chemical etching process. As shown in Fig.\ref{fig:8}(d), enormous nanopores are observed, which means the direct ion irradiation is a promising and feasible method for introducing nanopores into these various 2D materials. Among them, a large nanopore with a diameter more than 1.5 nm and a series of multiple tiny nanopores along the same line are observed as corresponding MD simulations, which demonstrates the proposed DP-Tab potental model could play a guiding role in the understanding of irradiation damage processes. Besides that, as the BCA schematic shown in Fig.\ref{fig:8}(a), the energy transferred from the initial ion to the primary knocked atom is $T=\dfrac{4M_1M_2}{(M_1+M_2)^2}E_1cos^2(\phi)$, here $E_1 = \dfrac{M_1}{2}\upsilon_1^2$ is the initial ion energy. To induce interlayer collision cascades, $\phi$ is close to $\dfrac{\pi}{2}$. That is to say, the ion energy determines the upper limitation of nanopores that could be generated under low dose irradiation. Therefore, controlling the ion energy could be an effective way to modulate nanopores' size. Note that 2D materials with tunable nanopores have a bright future in the applications including water desalination, molecular separation, DNA sequences and so on.\cite{RN610,RN611,RN612,RN613,RN614} More detail work about the nanopore formation in monolayer MoS$_2$ is undergoing. 

	Although the DP-Tab potential shows overall performance on the concerned material properties, the most challenge limits its future application in large systems is its computational cost. Previous studies has indicated that although the machine-learning potentials are generally more accurate than the classical analytical potentials, its computation cost is also 2 $\sim$ 4 orders of magnitude higher, which makes it challenging to obtain extensive statistics of MD simulations.\cite{RN354,RN228} However, since the tensor operations run much faster on the model GPU than the traditional CPU, we have tested the time spent to run one MD step using the DP-Tab, SNAP and SW-FM potential on the GPU node and the CPU node respectively. As shown in Fig.S3, the DP-Tab potential runs on the GPU node has obtained about 30-50 times speed-up compared to the CPU node and its speed is comparable to the SNAP potentials. According to the speed test result, we can infer that by using proper neural network parameters and choosing much faster Tensor Processing Unit (TPU)\cite{RN506} or Neural-network Processing Unit (NPU),\cite{RN495} it is promising that the speed of the DP-Tab potential could be even faster than the classical analytical SW-FM potential, which also means the application of deep-learning potential to large systems and long scales is practicable and foreseeable. Note that DPMD simulations with up to 100 million atoms have been achieved recently.\cite{RN573,RN574}

\section{Conclusion}
The all-electron calculations and the active-learning scheme are utilized to enhance the accuracy of a repulsive table potential modified deep-learning potential model for  irradiation damage simulations in monolayer MoS$_2$. The constructed potential could not only give an overall good performance on the predictions of near-equilibrium material properties including lattice constants, elastic coefficients, energy stress curves, phonon spectra, defect formation energy and displacement threshold energy and so on, but also reproduce the $ab$ $initial$ irradiation damage processes. Further irradiation simulations indicate that nanopores could be introduced into the monolayer MoS$_2$ directly by the single ion irradiation and sometimes a large nanopore with a diameter of more than 2 nm or a series of multiple nanopores could be generated by one single high-energy ion. These novel simulated phenomena are then qualitatively verified by subsequent 500 keV Au$^+$ ion irradiation experiments. Overall, our work provides a promising and feasible approach to explore the irradiation damage processes in enormous newly-discovered materials with unprecedented accuracy.


\begin{acknowledgments}
This work is supported by the Science Challenge Project (No. TZ2018004),    National Natural Science Foundation of China (Grant No. 91426304, 11705010 and 11871110), and the National Key Research and Development Program of China (Grants No. 2016YFB0201200 and 2016YFB0201203). We are grateful for computing resource provided by the High Performance Computing Platform of the Center for Life Science of Peking University, Weiming No. 1, Life Science No. 1 and Life Science Center High Performance Computing Platform at Peking University, the Terascale Infrastructure for Groundbreaking Research in Science and Engineering (TIGRESS) High Performance Computing Center and Visualization Laboratory at Princeton University, as well as TianHe-1(A) at National Supercomputer Center in Tianjin.
\end{acknowledgments}

\bibliography{MoS2_dptab_article}

\end{document}